# Phase and amplitude response of "0.7 feature" caused by holes in silicon one-dimensional wires and rings

**N T Bagraev[1], N G Galkin[1], W Gehlhoff[2], L E Klyachkin[1], and A M Malyarenko[1]**
[1]Ioffe Physico-Technical Institute, RAS, 194021 St. Petersburg, Russia
[2]Institut für Festkörperphysik, TU Berlin, D-10623 Berlin, Germany

E-mail: impurity.dipole@mail.ioffe.ru

**Abstract.** We present the findings of the $0.7 \cdot (2e^2/h)$ feature in the hole quantum conductance staircase that is caused by silicon one-dimensional channels prepared by the split-gate method inside the p-type silicon quantum well (SQW) on the n-type Si (100) surface. Firstly, the interplay of the spin depolarisation with the evolution of the $0.7 \cdot (2e^2/h)$ feature from the $e^2/h$ to $3/2\, e^2/h$ values as a function of the sheet density of holes is revealed by the quantum point contact connecting two 2D reservoirs in the p-type SQW. The 1D holes are demonstrated to be spin-polarised at low sheet density, because the $0.7 \cdot (2e^2/h)$ feature is close to the value of $0.5 \cdot (2e^2/h)$ that indicates the spin degeneracy lifting for the first step of the quantum conductance staircase. The $0.7 \cdot (2e^2/h)$ feature is found to take however the value of $0.75 \cdot (2e^2/h)$ when the sheet density increases thereby giving rise to the spin depolarisation of the 1D holes. Secondly, the amplitude and phase sensitivity of the $0.7 \cdot (2e^2/h)$ feature are studied by varying the value of the external magnetic field and the top gate voltage that are applied perpendicularly to the plane of the double-slit ring embedded in the p-type SQW, with the extra quantum point contact inserted in the one of its arms. The Aharonov-Bohm (AB) and the Aharonov-Casher (AC) conductance oscillations obtained are evidence of the interplay of the spontaneous spin polarisation and the Rashba spin-orbit interaction (SOI) in the formation of the $0.7 \cdot (2e^2/h)$ feature. Finally, the variations of the $0.7 \cdot (2e^2/h)$ feature caused by the Rashba SOI are found to take in the fractional form with both the plateaux and steps as a function of the top gate voltage.

## 1. Introduction
Progress in semiconductor nanotechnology makes it possible to fabricate clean one-dimensional (1D) constrictions with low density of high-mobility charge carriers, which exhibit ballistic behavior if the mean free path is longer than the channel length [1-8]. Therefore, the conductance of such quantum wires prepared by the split-gate [1-7] and cleaved edge overgrowth [8] methods depends only on the transmission coefficient, $T$ [9, 10]:

$$G_0 = g_s \frac{e^2}{h} N \cdot T \qquad (1)$$

where $N$ denotes the number of the highest occupied 1D subband, which is changed by varying the split-gate voltage, $U_{sg}$. Furthermore, the dependence $G(U_{sg})$ represents the quantum conductance staircase, because the conductance of a quantum wire is changed by the value of $g_s e^2/h$ each time when the Fermi level coincides with one of the 1D subbands [2, 3]. Spin factor, $g_s$, describes the spin degeneration of the wire mode. The value of $g_s$ is equal to two for non-interacting fermions if the external magnetic field is absent and becomes unity as a result of the Zeeman splitting of a quantum staircase in strong magnetic fields. The first step of the quantum conductance staircase has been found however to split into two parts even in the absence of external magnetic field [4-7]. The height of the substep that is dependent on temperature is usually observed to be about 0.7 of the first step value in a zero magnetic field. Two experimental observations indicate the importance of the spin component for the behavior of this $0.7 \cdot (2e^2/h)$ feature. First, the electron $g$ - factor was found to increase from 0.4 to 1.3 as the number of occupied 1D subband decreases [4]. Second, the height of the $0.7 \cdot (2e^2/h)$ feature attains to a value of $0.5 \cdot (2e^2/h)$ with increasing external magnetic field [4-7]. These results have defined the spontaneous spin polarisation of a 1D gas in a zero magnetic field as one of possible mechanisms for the *0.7·(2e²/h)* feature [7, 11-15] in spite of the theoretical prediction of a ferromagnetic state instability in ideal 1D systems in the absence of a magnetic field [16].

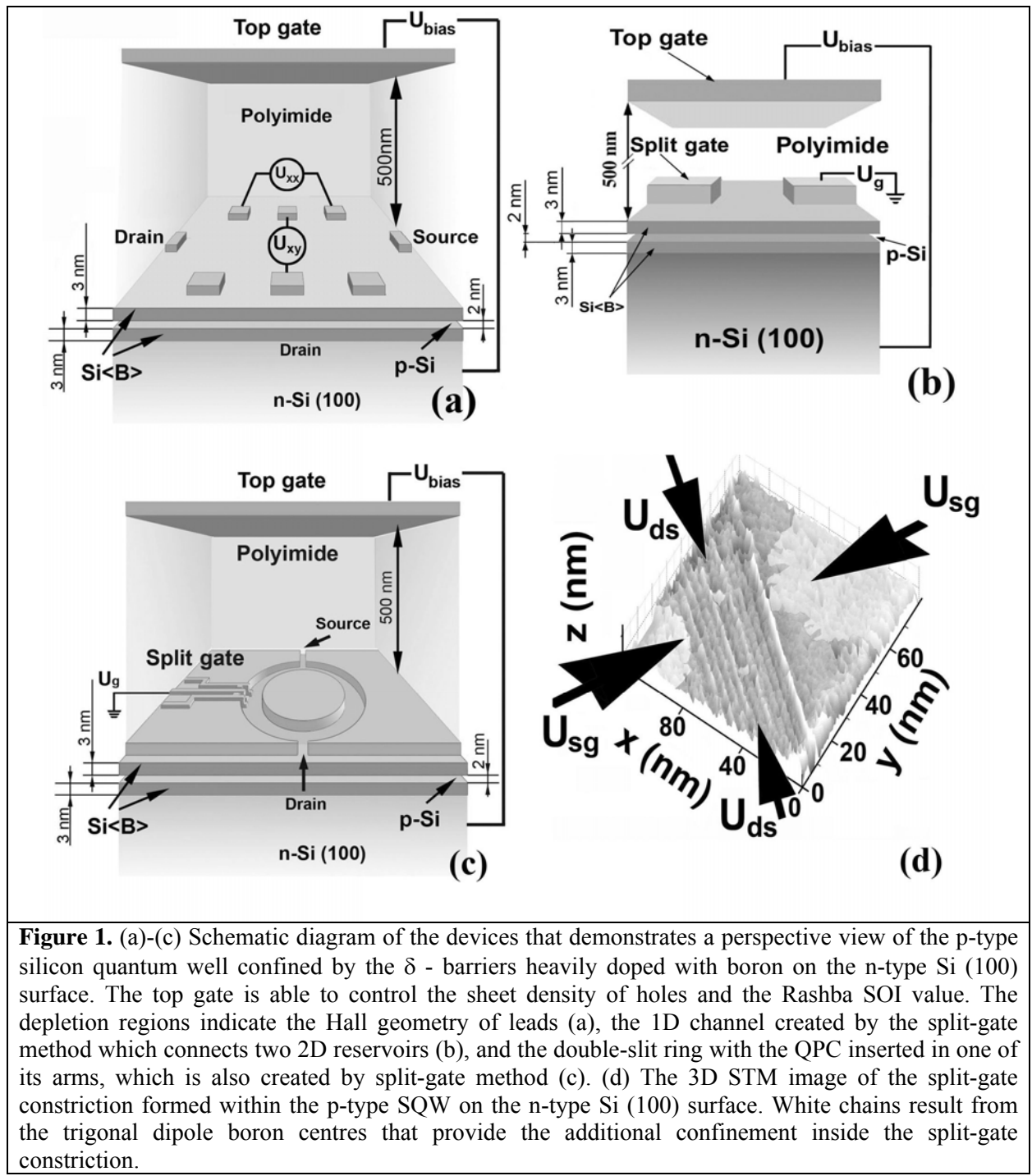

**Figure 1.** (a)-(c) Schematic diagram of the devices that demonstrates a perspective view of the p-type silicon quantum well confined by the δ - barriers heavily doped with boron on the n-type Si (100) surface. The top gate is able to control the sheet density of holes and the Rashba SOI value. The depletion regions indicate the Hall geometry of leads (a), the 1D channel created by the split-gate method which connects two 2D reservoirs (b), and the double-slit ring with the QPC inserted in one of its arms, which is also created by split-gate method (c). (d) The 3D STM image of the split-gate constriction formed within the p-type SQW on the n-type Si (100) surface. White chains result from the trigonal dipole boron centres that provide the additional confinement inside the split-gate constriction.

Studies of the quantum conductance staircase revealed by ballistic channels have shown that the *0.7·(2e²/h)* feature is observed not only in various types of the electron and hole GaAs based quantum wires [2-8, 11-13, 15, 17], but also in the hole Si based quantum wires [14, 18-20]. The latter findings were made possible by the developments of the diffusion nanotechnology that allows the fabrication of the p-type ultra-narrow silicon quantum wells (SQW) on the n-type Si (100) surface, which are confined by the δ - barriers heavily doped with boron (see figure 1a) [18, 21, 22]. Here these silicon quantum wires prepared by the split-gate method inside the p-type SQW are used to verify the interplay between the amplitude of the *0.7·(2e²/h)* feature and the degree of the spontaneous spin polarization that has been predicted in frameworks of the Kohn-Sham mean-field and the Hartree-Fock approximation with ultra-low linear concentration of charge carriers when the energy of the exchange interaction begins to exceed the kinetic energy in a zero magnetic field [23-25, 14]. The evolution of the *0.7·(2e²/h)* feature in the quantum conductance staircase from the $e^2/h$ to $3/2\ e^2/h$ values which is found when the sheet density of holes increases by varying the top-gate voltage seems to be caused by the spin depolarization processes in spin-polarised one-dimensional channels (see figure 1b).

In addition to the effect on the value of the sheet density of holes, the variations in the top-gate voltage are able to result in the enhancement of the Rashba spin-orbit interaction (SOI) due to the structure inversion asymmetry in mesoscopic nanostructures thereby being conducive to the spin interference of the 1D holes in the quantum wire [26-28]. The spin interference caused by the Rashba SOI value appeared to give rise to the developments of spintronic devices based on the spin-interference phenomena that are able to demonstrate the characteristics of the spin field-effect transistor (FET) even without ferromagnetic electrodes and external magnetic field [29, 30]. For instance, the spin-interference device shown schematically in figure 1c represents the Aharonov-Bohm (AB) ring covered by the top gate electrode, which in addition to the geometrical Berry phase provides the phase shift between the transmission amplitudes for the particles moving in the clockwise and anticlockwise direction [30]. This transmission phase shift (TPS) seems to be revealed by the Aharonov-Casher (AC) conductance oscillations measured by varying the top gate voltage applied to the two-terminal device with the only drain and source constrictions [31]. However, the variations in the density of carriers that accompany the application of the top gate voltage are also able to give rise to the conductance oscillations that result from the variations in the value of the Fermi wave vector, although similar in appearance to the AC conductance oscillations [32, 33]. Therefore the three-terminal device with the quantum dot (QD) [34] or the quantum point contact (QPC) [35-38] inserted in one of the ring's arms using the split-gate technique could be more appropriate to divide the relative contribution of the AC effect in the conductance oscillations (see figure 1c).

Since the AB ring's conductance has to be oscillated with the periodicity of a flux quantum h/e when a variable magnetic field threads its inner core, these AB oscillations have been shown to be persisted, if the transport through the QD [34] or QPC [35-38] inserted is coherent. The TPS caused by the QD or QPC has been found to be equal to π in the absence of the spin polarisation of carriers [34, 35-38]. Besides, the TPS value in the range of the *0.7·(2e²/h)* feature of the quantum conductance staircase revealed by the QPC inserted appeared to be equal to π/2 thereby verifying the spin polarisation in the AB ring [19, 20]. This TPS is not accompanied by changes in the amplitude of the *0.7·(2e²/h)* feature, thus remaining unsolved the question on the relative contribution of the spontaneous spin polarization and the Rashba SOI to its creation. Therefore here the three-terminal device provided by the top gate is used to reveal the interplay of the amplitude and phase sensitivity of the *0.7·(2e²/h)* feature by tuning the Rashba SOI (figure 1c). The AC conductance oscillations measured by varying the top-gate voltage in the range corresponding to the stability of the mobility and the sheet density of holes are shown to define the relative contribution of the spontaneous spin polarisation [11-13, 36-38] and other mechanisms [39-41] to the formation of the *0.7·(2e²/h)* feature.

## 2. Methods

The devices are based on the ultra-narrow, 2 nm, p-type high-mobility silicon quantum well (SQW) confined by the δ - barriers heavily doped with boron on the n-type Si (100) surface (figures 1a, b and c). These p-type SQW are prepared on the n-type Si (100) wafers during preliminary oxidation and subsequent short-time diffusion of boron by the CVD method [18, 21, 22, 42].

The preparation of oxide overlayers on silicon monocrystalline surfaces is known to be favourable to the generation of the excess fluxes of self-interstitials and vacancies that exhibit the predominant crystallographic orientation along a <111> and <100> axis, respectively [22, 42]. In the initial stage of the oxidation, thin oxide overlayer produces excess self-interstitials that are able to create small microdefects, whereas oppositely directed fluxes of vacancies give rise to their annihilation. Since the points of outgoing self-interstitials and incoming vacancies appear to be defined by the positive and negative charge states of the reconstructed silicon dangling bond, the dimensions of small microdefects of the self-interstitials type near the Si (100) surface have to be restricted to 2 nm. Therefore, the distribution of the microdefects created at the initial stage of the oxidation seems to represent the fractal of the Sierpinski Gasket type with the built-in self-assembled SQW [18]. Although Si-QWs embedded in the fractal system of self-assembled microdefects are of interest to be used as a basis of optically and electrically active microcavities in optoelectronics and nanoelectronics, the presence of dangling bonds at the interfaces prevents such an application. Therefore, subsequent short-time diffusion of boron would be appropriate for the passivation of dangling bonds and other defects created during previous oxidation of the Si (100) surface thereby assisting the transformation of the arrays of microdefects in the neutral δ - barriers confining the ultra-narrow, 2nm, SQW (figures 1a, b and c).

We have prepared the p-type SQWs with different density of holes ($10^9 \div 10^{12}$ cm$^{-2}$) on the Si (100) wafers of the n-type in frameworks of the conception discussed above and identified the properties of the two-dimensional gas of holes by the cyclotron resonance (CR), Hall-effect and infrared Fourier and tunneling spectroscopy methods. The energy positions of two-dimensional subbands for the light and heavy holes in the SQW studied were determined by studying the far-infrared electroluminescence spectra obtained with the infrared Fourier spectrometer IFS-115 Brucker Physik AG as well as by using the local tunnelling spectroscopy technique [22, 38]. The results obtained are in a good agreement with corresponding calculations following by Ref [43] if the width of the SQW, 2nm, is taken into account. The secondary ion mass spectroscopy (SIMS) and scanning tunneling microscopy (STM) studies have shown that the δ - barriers, 3 nm, heavily doped with boron, 5 $10^{21}$ cm$^{-3}$, represent really alternating arrays of undoped and doped tetrahedral dots with dimensions restricted to 2 nm. The value of the boron concentration determined by the SIMS method seems to indicate that each doped dot located between undoped dots contains two impurity atoms of boron. Nevertheless, the angular dependencies of the cyclotron resonance spectra and the conductivity have demonstrated that these p-type SQW confined by the δ - barriers heavily doped with boron contain the high mobility 2D hole gas that is characterized by long transport relaxation time of heavy and light holes at 3.8 K, $\tau \geq 5 \cdot 10^{-10}$ c [21, 44, 45]. Thus, the transport relaxation time of holes in the ultra-narrow SQW appeared to be longer than in the best MOS structures contrary to what might be expected from strong scattering by the heavily doped δ - barriers. This passive role of the δ - barriers between which the SQW is formed was quite surprising, when one takes into account the high level of their boron doping. To eliminate this contradiction, the temperature dependencies of the conductivity and the Seebeck coefficient as well as the EPR spectra and the local tunneling current-voltage characteristics have been studied [18, 21]. The EPR and the thermo-emf studies show that the boron pairs inside the δ - barriers are the trigonal dipole centres, $B^+$-$B^-$, which are caused by the negative-U reconstruction of the shallow boron acceptors, $2B_0 \Rightarrow B^+ + B^-$. In common with the other solids that consist of small bipolarons, the δ - barriers containing the dipole boron centres have been found to be in an excitonic insulator regime at the sheet density of holes in the SQW lower than $10^{11}$ cm$^{-2}$. However, the electrical resistivity, thermo-emf and magnetic susceptibility measurements demonstrate that the high sheet density of holes in the SQW (>$10^{11}$ cm$^{-2}$) gives rise to the superconductor properties for the δ - barriers in frameworks of the mechanism of the single-hole tunneling through the negative-U centres which is in the interplay with the multiple Andreev reflections inside SQW [18].

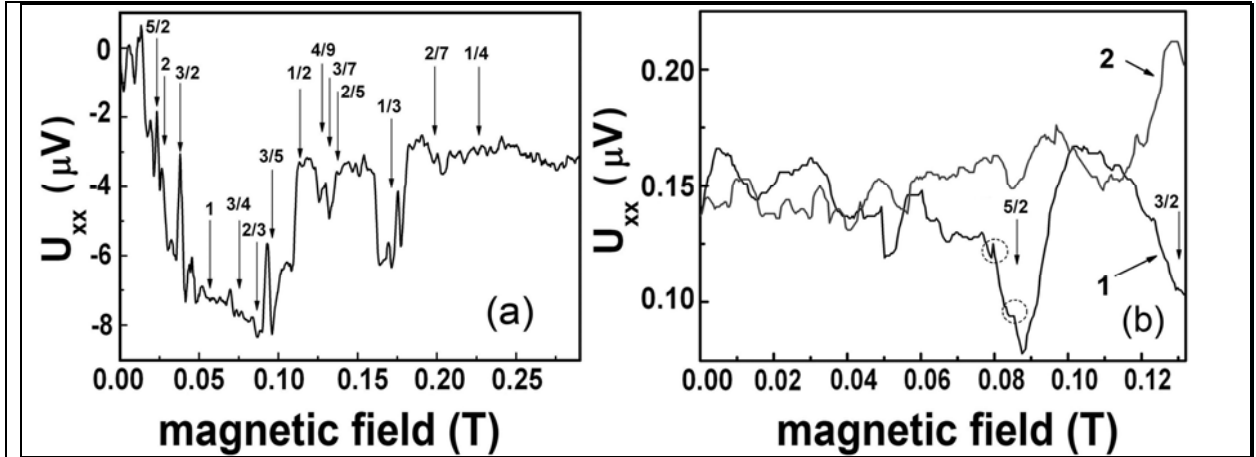

**Figure 2.** Shubnikov - de Haas oscillations in external magnetic field that is perpendicularly (a) and in parallel (b) with the SQW plane, T=77 K. (b) The curves 1 and 2 correspond to antiparallel orientation of the external magnetic field along the $U_{xx}$ axis (see figure 1a). (a) $p_{2D} = 3 \cdot 10^{13}$ m$^{-2}$; $m_{eff} = 5.2 \cdot 10^{-4}$ $m_0$. (b) $p_{2D} = 1.1 \cdot 10^{14}$ m$^{-2}$; $m_{eff} = 2.6 \cdot 10^{-4}$ $m_0$. Dashed circles depict the closest AAS features.

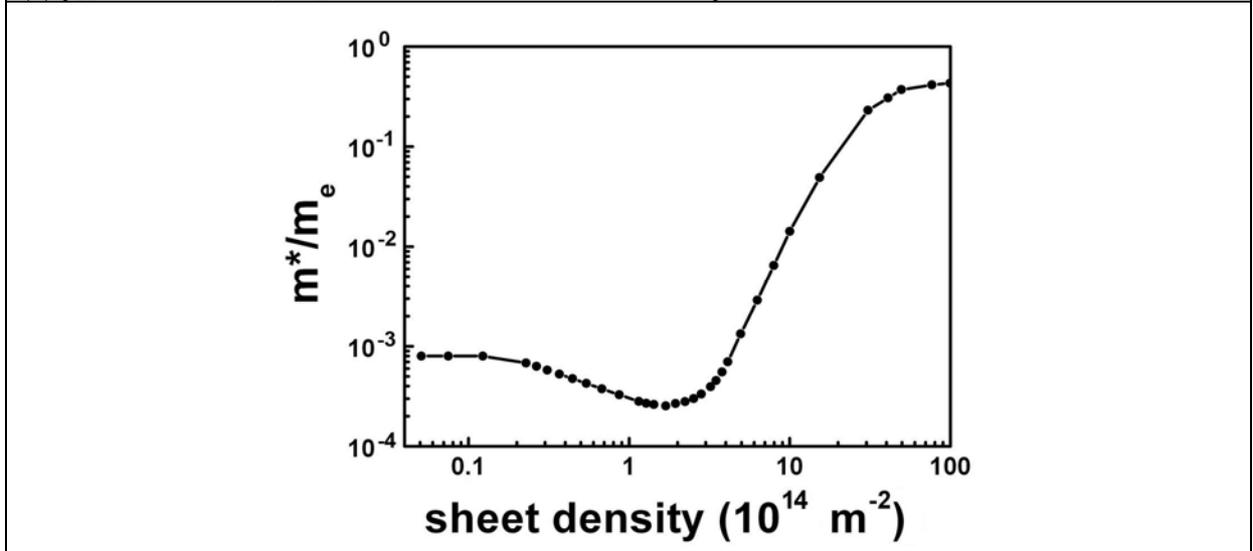

**Figure 3.** The effective mass of heavy holes in the type SQW confined by the δ - barriers heavily doped with boron on the n-type Si (100) surface versus the sheet density of holes.

These extraordinary properties of the p-type SQW confined by the δ - barriers heavily doped with boron emerge in the Shubnikov – de Haas (SdH) oscillations measured at high temperatures, 77 K (figures 2a and b). These findings became it possible owing to the small effective mass of the heavy holes that was controlled by studying the temperature dependences of the SdH oscillations in the low sheet density SQW (see figure 3). The effective mass values appear to be in a good agreement with the CR data and the estimations from the period of the AC oscillations that are given below.

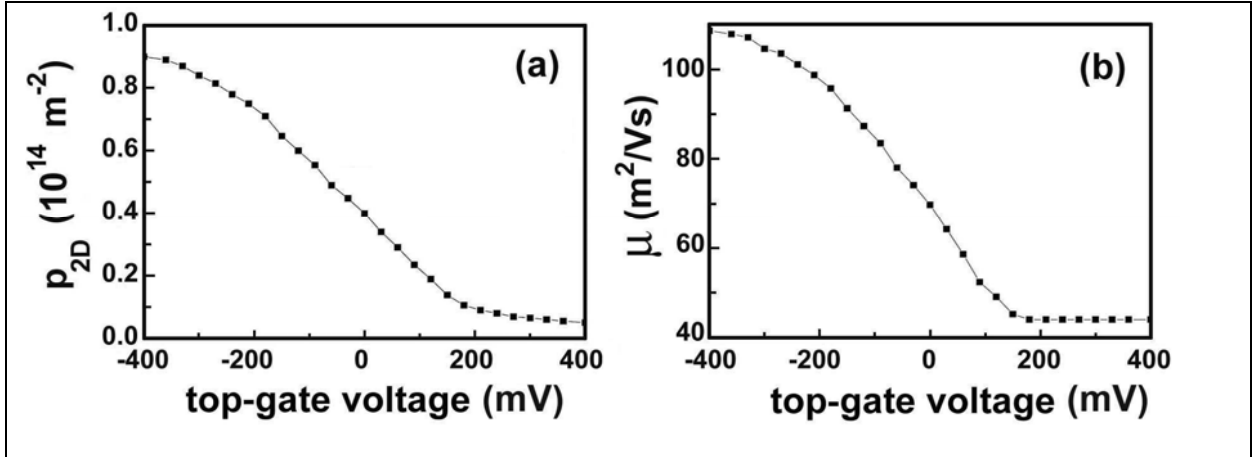

**Figure 4.** Sheet density (a) and mobility (b) revealed by the two-dimensional gas of holes as a function of the top gate voltage which were extracted from the longitudinal and the Hall voltage measurements of the p-type SQW confined by the δ-barriers on the n-type Si (100) surface.

The SdH oscillations are very surprisingly to be observed also in the external magnetic field that is in parallel with the SQW plane (figure 2b) along with the ordinary SdH (figure 2a). Since the emf, $U_{xx}$, of the opposite sign relatively to the direction of the longitudinal magnetic field is observed, the SdH oscillations shown in figure 2b seem to result from the diamagnetic response from the δ - barriers. This premise is supported by the presence of additional features superimposed on the oscillations which are caused by the Aharonov – Altshuler - Spivak (AAS) oscillations taking account of the SQW width, 2 nm, and the dimensions of the structure in figure 1a (see dashed circles in figure 2b) [46]. Thus, the preparation of the ultra-narrow p-type SQW confined by the δ - barriers which in properties and composition is quite similar to graphene [47] made it possible for the first time to use the split-gate constriction to study the quantum conductance staircase of holes at the temperature of 77 K [49-51]. Furthermore, even with small drain-source voltage the electrostatically ordered dipole centres of boron within the δ - barriers are able to stabilize the formation of the one-dimensional subbands, when the quantum wires are created inside SQW using the split-gate technique (see figure 1d) [21].

The devices applied here to analyse the dependence of the *0.7·(2e²/h)* feature on the sheet density of holes were prepared at the same SQW (see figures 1a, b and c). The parameters of the high mobility SQW were defined by the Hall measurements (see figures 4a and b). The initial value of the sheet density of 2D holes, $4·10^{13}$ m$^{-2}$, was changed controllably over one order of magnitude, between $5·10^{12}$ m$^{-2}$ and $9·10^{13}$ m$^{-2}$, by biasing the top gate above a layer of insulator, which fulfils the application of the p$^+$-n bias junction. The variations in the mobility measured at 3.8 K that corresponded to this range of the $p_{2D}$ values appeared to occur between 80 and 420 m²/vs. Thus the value of the mobility was high even at low sheet density. Besides, the high value of mobility showed a decrease no more than two times in the range of temperatures from 3.8 K to 77 K that seems to be caused by the extraordinary properties for the δ – barriers that have been described above. These characteristics of the 2D gas of holes allowed the studies of the quantum conductance staircase revealed by the heavy holes at 77 K. The number of the highest occupied mode of the quantum wire was controlled by varying the split-gate voltage, whereas the sheet density of holes and the Rashba SOI was tuned by biasing the top gate (see figures 1b and c). The experiments are provided by the effective length of the quantum wire, 0.2 μm, and the cross section of the 1D channel, 2 nm × 2 nm, which is determined by the width of the SQW and the lateral confinement due to the properties for the δ – barriers (figure 1b).

The one-dimensional ring embedded electrostatically in the SQW, R=2500 nm, contains the source and the drain constrictions that represent quantum point contacts (QPCs) as well as the QPC inserted in its arm by the split-gate method (see figure 1c). The effective length of the QPC inserted is equal also to 0.2 μm; the cross section of the 1D ring is 2 nm × 2 nm, which is also determined by the width of the SQW and the lateral confinement due to the properties for the δ – barriers (figure 1c).

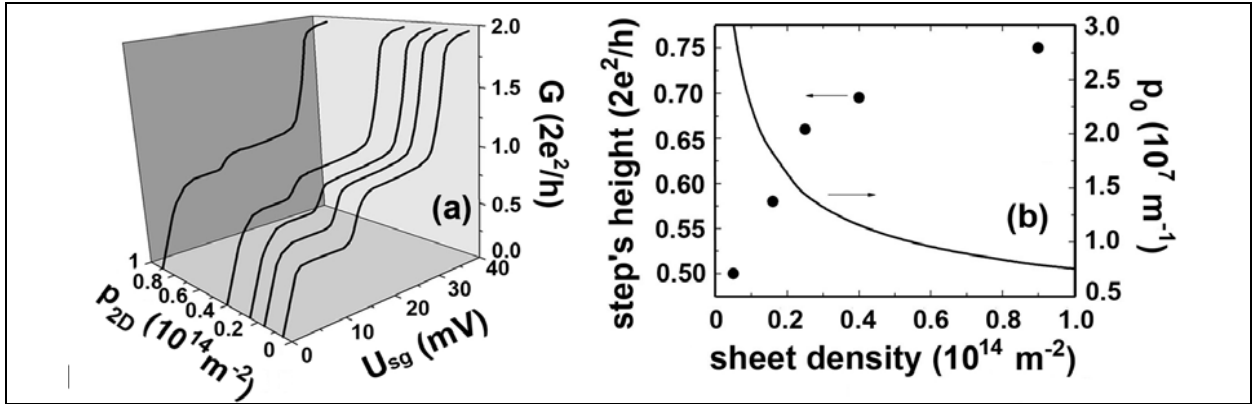

**Figure 5.** (a) The quantum conductance staircase of the silicon quantum wire as a function of the sheet density of holes that was tuned controllably by biasing the top gate, which fulfils the application of the $p^+$-$n$ bias voltage to the p-type silicon quantum well on the n-type Si (100) surface, T=77 K. The $p^+$-$n$ bias voltage is varied from the forward branch to the reverse branch between +110 mV and -120 mV, which establishes the range of the magnitude of $p_{2D}$ from $5\cdot10^{12}$ m$^{-2}$ to $9\cdot10^{13}$ m$^{-2}$ provided that zero $p^+$-$n$ bias voltage results in the value of $p_{2D}$ equal to $4\cdot10^{13}$ m$^{-2}$.
(b) Dependence of the critical linear concentration corresponding to a complete spin depolarization of the quasi-1D electron gas in a quantum wire connecting 2D reservoirs in a silicon quantum well on the sheet density of holes; the cross section of the 1D channel is 2 nm × 2 nm. Circles indicate the values of the height of the $0.7\cdot(2e^2/h)$ feature that are shown in figure 5a.

### 3. Results

*3.1. Spin depolarisation and quenching of the $0.7\cdot(2e^2/h)$ feature in the quantum conductance staircase of a quantum wire*

Figure 5a shows the quantum staircase revealed by the heavy holes in the 1D channel defined by the split-gate voltage inside the SQW at different sheet density of holes. The $0.7\cdot(2e^2/h)$ feature is seen to be coincident practically with its normal value provided that the top-gate voltage is kept to be zero, which appeared to result in the value of $p_{2D}$ equal to $4\cdot10^{13}$ m$^{-2}$. Tuning the value of $p_{2D}$ by biasing the top gate causes however the variations in the height of this feature. The value of $0.5\cdot(2e^2/h)$ is found under the forward branch of the top-gate voltage, whereas at large values of $p_{2D}$ induced by the reverse branch of the top-gate voltage the $0.7\cdot(2e^2/h)$ feature attains the value of $0.75\cdot(2e^2/h)$ (see figure 4a). The height of the $0.7\cdot(2e^2/h)$ feature studied as a function of the hole concentration in the p-type silicon SQW is worthwhile to be related to the behavior of the critical linear concentration, $p_0$, with the exchange energy that compensates the kinetic energy, which was calculated by extrapolation from the known dependence of the hole effective mass in the p-type silicon quantum wells on the value of $p_{2D}$ [14], (see figures 3 and 5b).
If the linear concentration $p_{1D}$ is less than a critical value $p_0$, the energy of the exchange interaction exceeds the kinetic energy and, thus, the spin-polarised state is energetically more favourable than the unpolarised state. At the same time, if the linear concentration of holes exceeds the critical value, $p_0$, and the kinetic energy is dominant, the unpolarised state becomes to be more favourable, with the value of $p_0$ dependent only on the width of the quantum wire and the effective mass of charge carriers [14]. It should be noted that the exchange interaction may significantly affect the effective mass of charge carriers in quantum wires, because the spin-polarised states with extended wave functions in a quasi-1D system are spontaneously formed at higher values of $p_{2D}$ than in a 2D gas [24]. Therefore the kinetic energy is effectively quenched in the middle part of a quantum wire that connects two 2D reservoirs, because a competition with the exchange energy is available, which may favour a reduction in the effective mass with increasing the values of $p_{2D}$. Such a gain in the exchange interaction may account for a rise in the effective mass of holes as the their sheet density decreases extremely in the SQW (see figure 3), because the low density 2D gas is able to decay in the system of two-dimensional lakes connected by quantum wires or quantum point contacts, which result from specifically in the presence of disorder [48].

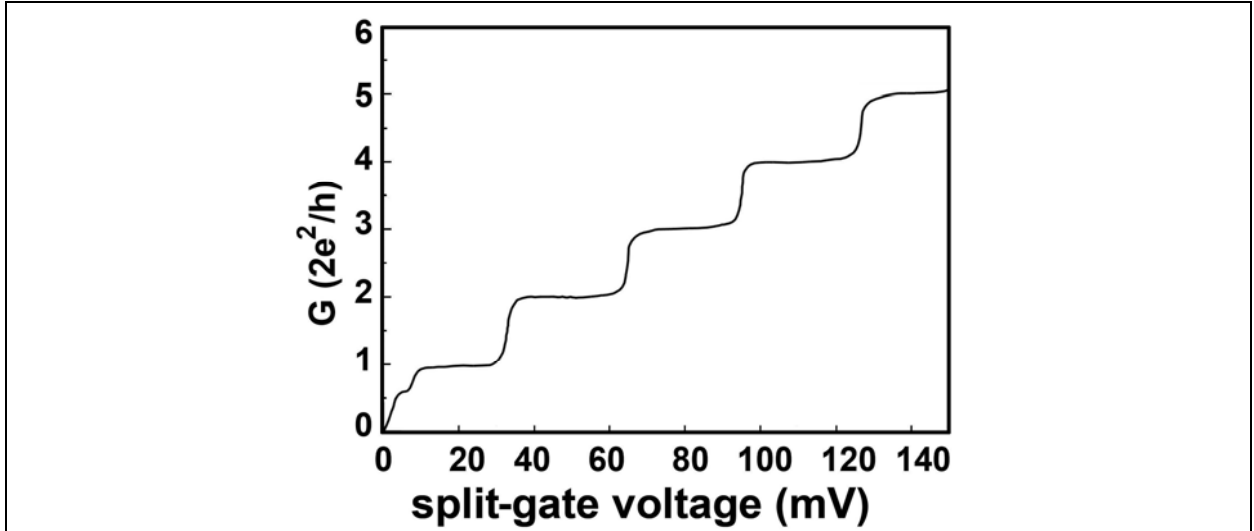

**Figure 6.** The hole quantum conductance staircase as a function of the split-gate voltage which is revealed by the QPC inserted in one of the arms of the Si-based double-slit ring in zero magnetic field and under zero bias voltage controlled by the top gate (see figure 1c).

Thus, the variations in the height of the *0.7·(2e²/h)* feature that result from controllable tuning the value of $p_{2D}$ can be perceived as a result of partial spin depolarization of holes, which is enhanced as the critical linear concentration in the 1D channel, $p_0$, is approached. Finally, the behavior of the *0.7·(2e²/h)* feature in the quantum conductance staircase shares a common trait related to the critical linear concentration of holes and electrons that corresponds to their complete spin depolarization in the 1D channels prepared respectively inside the p-type Si based quantum well studied in this work and inside the n-type GaAs based quantum well discussed above [7].

*3.2. Fractional form of the 0.7·(2e²/h) feature in the quantum conductance staircase revealed by the quantum point contact inserted in one of the arms of a double-slit ring*

Figure 6 demonstrates the quantum conductance staircase revealed by the QPC inserted in one of the arms of the double-slit ring described above (see figure 1c). It should be noted that the quality of the quantum conductance staircase has been provided additionally by the electrostatic ordering of the dipole boron centres which results in the strong confinement along one-dimensional ring [18, 21, 49-51]. The presence of the pronounced steps made it possible to reproduce the results on the additional TPS in the range of the *0.7·(2e²/h)* feature found by varying the split-gate voltage firstly in the studies of the self-assembled silicon one-dimensional rings [19, 20]. This additional TPS revealed by the period mismatch of the AB conductance oscillations appeared to be equal to $\pi/2$ that is evidence of the spin polarization of heavy holes. Besides, the change of the TPS value from $\pi/2$ to $\pi$ under saturation of the electrically-detected NMR of the $^{29}$Si nuclei was strong verification of the spin polarization in the 1D channel (see also [19, 20, 51]). But, the TPS is not accompanied by the changes in the amplitude of the $0.7\cdot(2e^2/h)$ feature. Therefore, as noticed above, the variations in the top-gate voltage are required to define the relative contribution of the spontaneous spin polarization and the Rashba SOI to the mechanism of the $0.7\cdot(2e^2/h)$ feature. Here we focus on the effects of the bias voltage controlled by the top gate and the external magnetic field on the amplitude and the TPS of the $0.7\cdot(2e^2/h)$ feature at the fixed value of the split-gate voltage, 5.7 mV, in its range (see figure 6).

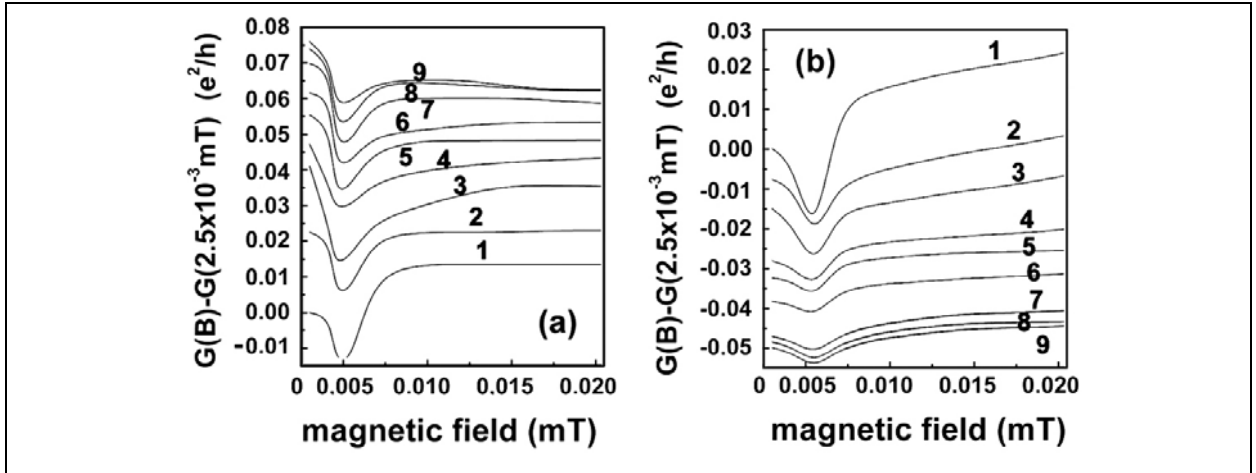

**Figure 7.** Experimental magnetoconductance, $G(B)-G(2.5\cdot 10^{-3} mT)$, for different negative (a) and positive (b) $U_{bias}$ determined by the top gate voltage. At zero magnetic field and top-gate voltage, the conductance was fixed by the split-gate voltage at 5.7 mV in the range of the $0.7\cdot(2e^2/h)$ feature. T=77 K. $B=2.5\cdot 10^{-3}$ mT is the residual magnetic field obtained after screening the Earth magnetic field. A crossover from the weak localisation to the weak antilocalisation is revealed by varying the top-gate voltage from negative to positive $U_{bias}$.
(a) $U_{bias}$, mV: 1 – 40, 2 – 140, 3 – 230, 4 – 260, 5 – 300, 6 – 330, 7 – 370, 8 – 390, 9 – 400.
(b) $U_{bias}$, mV: 1 – 30, 2 – 90, 3 – 140, 4 – 230, 5 – 260, 6 – 300, 7 – 360, 8 – 370, 9 – 380.

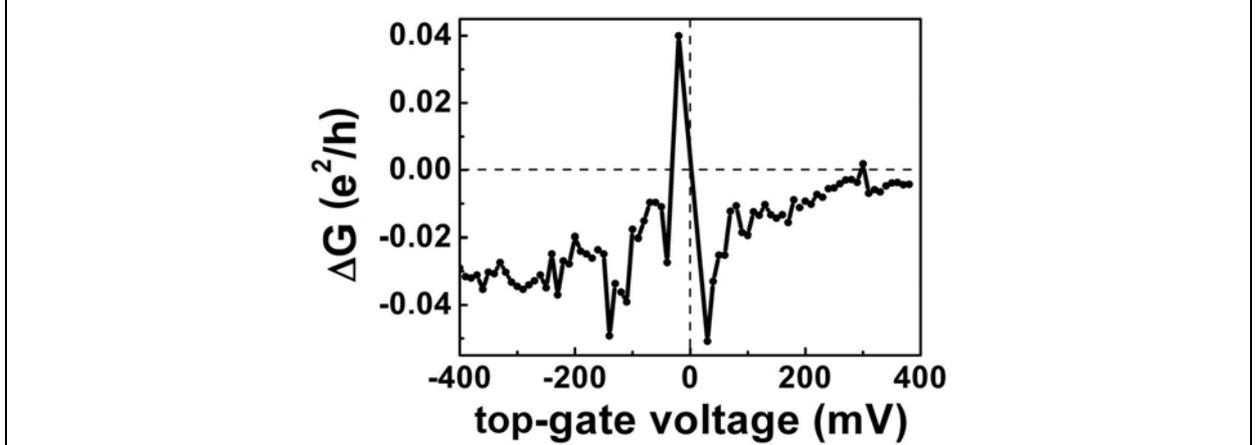

**Figure 8.** The magnetoconductance, $G=G(B)-G(0)$, as a function of the top-gate voltage applied to the gate-controlled double-slit ring with the short quantum wire inserted in one of its arms. At zero magnetic field and top-gate voltage, the conductance was fixed by the split-gate voltage at 5.7 mV in the range of the $0.7\cdot(2e^2/h)$ feature. B=0.0055 mT. T=77 K.

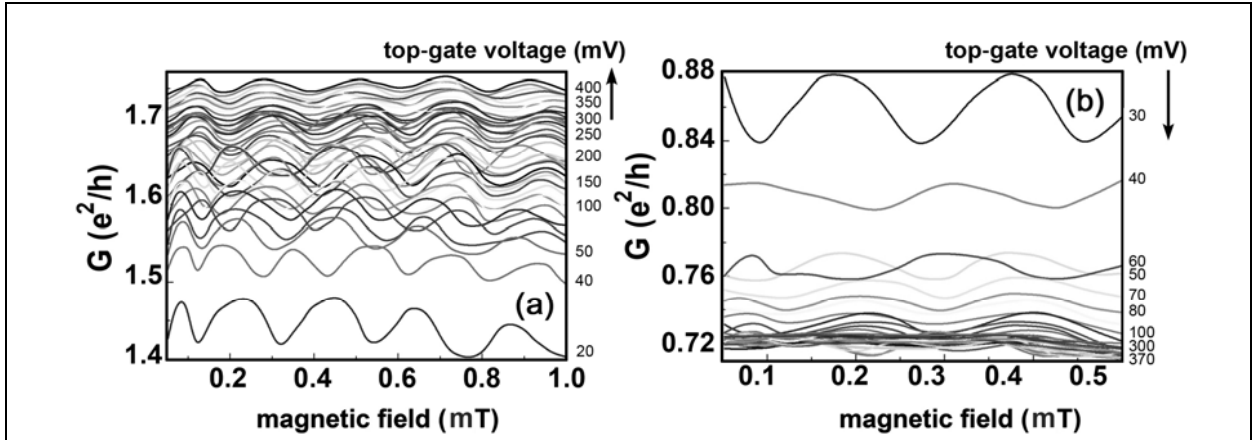

**Figure 9.** The Aharonov-Bohm conductance oscillations that attend the changes in the amplitude and the phase of the *0.7·(2e²/h)* feature under negative (a) and positive (b) top-gate voltage applied to the gate-controlled double-slit ring with the QPC inserted in one of its arms. At zero magnetic field and top-gate voltage, the conductance was fixed by the split-gate voltage at 5.7 mV in the range of the *0.7·(2e²/h)* feature. T=77 K.

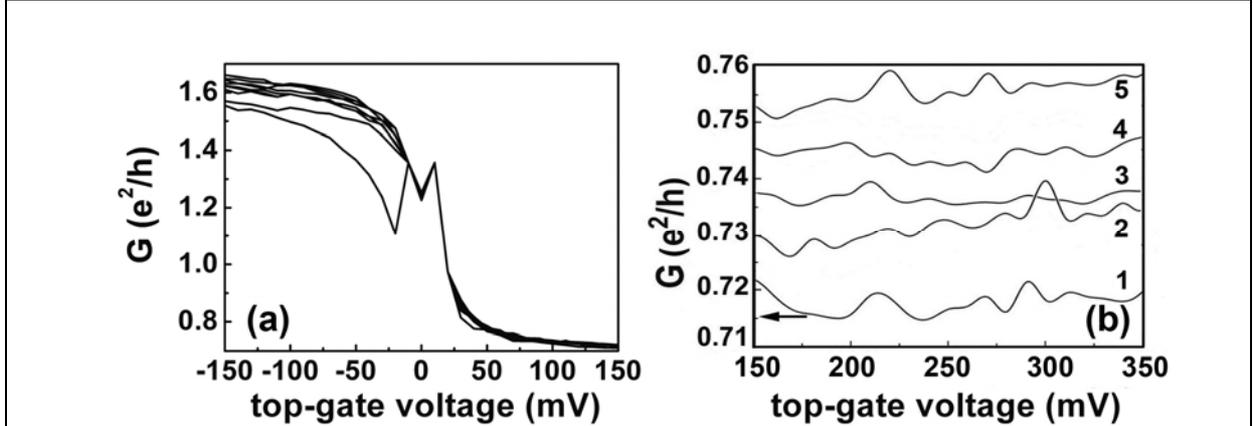

**Figure 10.** (a) The variations in the amplitude of the *0.7·(2e²/h)* feature as a function of the top-gate voltage at different magnetic fields in the double-slit ring with QPC inserted in one of its arms. (b) The AC conductance oscillations in the double-slit ring with QPC inserted in one of its arms. The external magnetic field value was changed from 0.05 mT (curve 1) to 0.45 mT (curve 5) in 0.1 mT step. The curves 2÷5 are displaced from the curve 1 an amount (0.00775 $e^2/h$)·N; where N is the curve number.

The presence of the Rashba SOI in the device studied is revealed by the measurements of the magnetoconductance (figures 7a and 7b). By varying the top gate voltage at the fixed value of the split-gate voltage, 5.7 mV, the transition from the positive magnetoresistance to the negative magnetoresistance is observed thereby verifying a crossover from the weak antilocalization to the weak localization following the changes of the concentration of the 2D holes. The dependencies shown in figures 7a and 7b are taken into account to be in a good agreement with the theoretical predictions and correlate with the experimental magnetoconductance data in a high mobility electron $In_xGa_{1-x}As/InP$ quantum well [52, 53]. Nevertheless, the positive component of magnetoresistance increases again in the range of the extremely low sheet density of holes that seems to result from the behaviour of the effective mass of the heavy holes in frameworks of the spontaneous spin polarisation (see figure 3). The experimental verification of this suggestion is possible to be followed by the analysis of the dependence of the magnetoconductance on the top-gate voltage value that appears to reveal the spin splitting in the 1D gas (see figure 8). The value of the spin splitting, 0.044 eV, is of interest to be coincident with the superconducting gap value revealed by the multiple Andreev reflections from the δ - barriers [22].

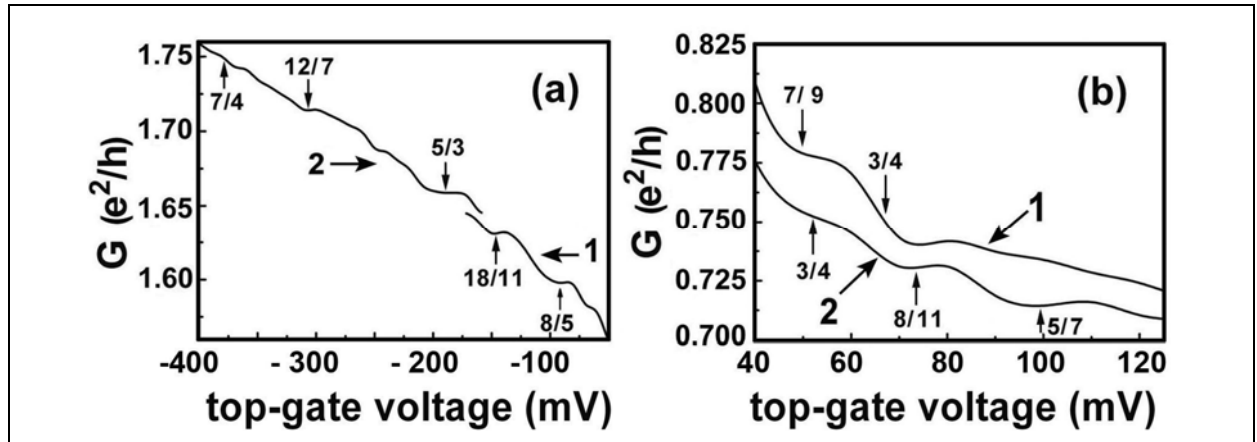

**Figure 11.** (a), (b) The fractional form of the *0.7·(2e²/h)* feature that are revealed by varying the value of the Rashba SOI controlled by varying the top-gate voltage.
B: (a) 1-0.157 mT; 2-0.253 mT; (b) 1- 0.0025 mT; 2-B=0.0080 mT;

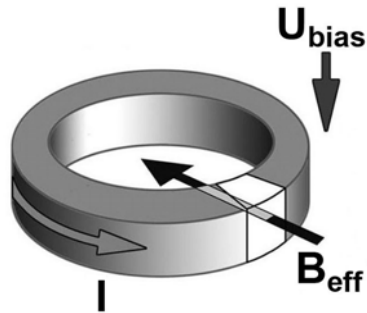

**Figure 12.** The scheme of the AB ring with inserted quantum point contact. $B_{eff}$ is the effective magnetic field induced by the Rashba SOI.

Thus, the Rashba SOI effects observed in the studies of the magnetoconductance are evidence of the constructive and destructive backscattering associated with QPCs by varying the top-gate voltage thereby allowing the findings of the AB conductance oscillations in the range of the external magnetic fields outside the region of the weak antilocalisation (figures 9a and b). The AB conductance oscillations observed are seen to exhibit the changes in both the phase and amplitude as a result of the variations in the top-gate voltage. Therefore these changes are successfully revealed by the corresponding dependences of the *0.7·(2e²/h)* feature value fixed at $U_{sg}$ = 5.7 mV as a function of the top-gate voltage at different magnetic fields.

Figure 10a shows the changes in the amplitude of the *0.7·(2e²/h)* feature found by biasing the top gate that appeared to be followed by increasing and decreasing the $p_{2D}$ value from the initial value, $4·10^{13}$ m$^{-2}$, with applying respectively the reversal and forward bias. The amplitude of the *0.7·(2e²/h)* feature reduces under the forward bias crossing the value of *0.5·(2e²/h)* that indicates the spin degeneracy lifting for the first step of the quantum conductance staircase [2]. The value of the spin splitting, 0.044 eV, which is revealed in weak magnetic fields and quite similar to that observed in 2D electron gas [12] appears to be the same as found by studying the magnetoconductance in the weak antilocalisation regime. These data seem to be evidence of the spontaneous spin polarization of heavy holes in the 1D channel that is due to the efficient quenching of the kinetic energy by the exchange energy of carriers [14]. The amplitude of the *0.7·(2e²/h)* feature fixed by the split-gate voltage is also found to exhibit the AC conductance oscillations in the absence of changes in the $p_{2D}$ and μ values that are revealed specifically by varying the forward bias voltage applied to the top gate, (see figures 10b, 4a and 4b). Indeed, the only range of the top-gate voltage between 150 and 400 mV can be used to divorce the AC conductance oscillations caused by the Rashba SOI from the other effects such as the Fermi resonances that give rise to the conductance oscillations by varying the sheet density and the mobility of 2D holes (figures 4a and 4b). These phase variations of the AC oscillations appear to be caused by the elastic scattering of the heavy holes on the QPCs inside the double-slit ring. The phase shifts calculated in frameworks of this model appeared to be dependent on the Rashba parameter, α, determined by the effective magnetic field, which is created by the Rashba SOI [26-33, 36, 54],

$$\mathbf{B}_{eff} = \frac{\alpha}{g_B \mu_B}[\mathbf{k} \times \mathbf{e}_z] \tag{2}$$

and consequently on the modulation of the conductance [27, 28], which is found to be in a good agreement with the AC conductance oscillations shown in figure 10b. It should be noted that the Rahba SOI in the 2D systems depends cubically on the wavevector. However, in the 1D system one should average the components of the wavevector perpendicular to the quantum wire axis. Thus, the resulting Rashba SOI for the 1D systems of holes depends linearly on the wavenumber as the Dresselhaus SOI in the 1D systems of electrons.

The period of the AC conductance oscillations revealed by holes in the QPC can be estimated from the relationships introduced by Winkler et al [27, 28]:

$$\Delta V_g \approx \frac{\hbar^2 d^2 l}{3\pi^2 R m_{eff} \beta_{hh}} \tag{3}$$

Where *l* is a characteristic length, which gives the proportionality between the top gate voltage $V_g$ and the electric field, $V_g = E_z l$, and is determined by the thickness of the n-type Si (100) wafer, *l*=300 μm; *R* is the radius of the ring, 2500 nm; *d* is the diameter of the quantum wire, 2nm. The value of $\beta_{hh}$ is caused by the Rashba parameter, α:

$$\alpha_{hh} = -3\beta_{hh}\langle k_r^2 \rangle E_z \tag{4}$$

$$\beta_{hh} = a(\gamma_2 + \gamma_3)\gamma_3 \left[ \frac{1}{\varepsilon_1^{hh} - \varepsilon_1^{lh}}\left(\frac{1}{\varepsilon_1^{hh} - \varepsilon_2^{lh}} - \frac{1}{\varepsilon_1^{hh} - \varepsilon_2^{hh}}\right) + \frac{1}{(\varepsilon_1^{hh} - \varepsilon_2^{lh})(\varepsilon_1^{hh} - \varepsilon_2^{hh})} \right]\frac{e\hbar^4}{m_{eff}^2} \approx$$

$$\approx 0.02 \left[ \frac{1}{\varepsilon_1^{hh} - \varepsilon_1^{lh}}\left(\frac{1}{\varepsilon_1^{hh} - \varepsilon_2^{lh}} - \frac{1}{\varepsilon_1^{hh} - \varepsilon_2^{hh}}\right) + \frac{1}{(\varepsilon_1^{hh} - \varepsilon_2^{lh})(\varepsilon_1^{hh} - \varepsilon_2^{hh})} \right]\frac{e\hbar^4}{m_{eff}^2} \tag{5}$$

where the value of $\langle k_r^2 \rangle$ is easily estimated as $\langle k_r^2 \rangle \approx \pi^2/d^2$; *d* is the diameter of the quantum wire; $a = \frac{64}{9\pi^2} \approx 0.7$ [28]; $\gamma_2 = -0.18$, $\gamma_3 = -0.1$ are the Luttinger parameters in Silicon [55]; $\varepsilon_{1,2}^{lh,hh}$ are the energies of the light and heavy holes in the SQW; where the lower index shows the number of the subband, whereas the upper index corresponds to the light or heavy hole. The optical studies allowed the energies of both light and heavy holes in the SQW used in the device: $\varepsilon_1^{hh} = 90 meV$; $\varepsilon_1^{lh} = 114 meV$; $\varepsilon_2^{hh} = 307 meV$; $\varepsilon_2^{lh} = 476 meV$.

The value of the effective mass of the heavy holes, $m_{eff}$, estimated from the period of the AC conductance oscillations as $6.7 \cdot 10^{-4} m_0$ is in a good agreement with the data of the CR measurements and the temperature dependences of the SdH oscillations. Owing to such small value of $m_{eff}$ that is due to the properties of the δ - barriers and the presence of the ultra-shallow $p^+$-*n* bias junction, these measurements of the spin interference have been realised at T=77 K.

The variations of the top gate voltage are very surprisingly to give rise to the fractional form of the $0.7\cdot(2e^2/h)$ feature with both the plateaux and the steps as a function of the top-gate voltage (see figures 11a and b). The plateaus and steps observed are of interest to bring into correlation with the odd and even fractions that seem to be caused by the geometry of the three-terminal device which is close to the diagram of the metallic loop suggested by R B Laughlin for the explanation of the quantized Hall conductivity (figure 12) [56]. In frameworks of this model, the external magnetic is replaced by the effective magnetic field that results from the Rashba SOI which effects only on the spin-dependent part of the one-dimensional transport. Therefore the processes of the spin-dependent scattering related to the multiple Andreev reflections from the δ – barriers containing the dipole boron centres seem to of importance for the conservation of spin polarisation in QPCs.

## 4. Conclusions

The dependence of the critical linear concentration that defines a complete spin depolarisation in a 1D channel connecting two 2D reservoirs on the sheet density of holes has been derived to analyse corresponding the evolution of the $0.7\cdot(2e^2/h)$ feature from the $0.5\cdot(2e^2/h)$ to $0.75\cdot(2e^2/h)$ values in the quantum conductance staircase revealed by the quantum wire prepared by the split-gate method inside the p-type ultra-narrow silicon quantum well on the Si (100) surface. The 1D channel studied seems to be spin-polarised at the linear concentration of holes lower than the critical linear concentration, because the $0.7\cdot(2e^2/h)$ feature is close to the value of $0.5\cdot(2e^2/h)$ that indicates the spin degeneracy lifting for the first step of the quantum conductance staircase. The $0.7\cdot(2e^2/h)$ feature has been found however to tend to the value of $0.75\cdot(2e^2/h)$ when the linear concentration of holes attains the critical value corresponding to the spin depolarisation.

The amplitude and the phase of the $0.7\cdot(2e^2/h)$ feature of the hole quantum conductance staircase have been shown experimentally to be controlled by tuning the Rashba coupling parameter induced by the top-gate voltage applied to the Aharonov - Bohm silicon ring with the quantum point contact inserted in one of its arms. The Aharonov - Casher conductance oscillations have been observed as the variations of the amplitude of the $0.7\cdot(2e^2/h)$ feature in the absence of the changes in the mobility and the sheet density of 2D holes in the p-type ultra-narrow silicon quantum well. The interplay between the Rashba SOI and spontaneous spin polarization of holes that gives rise to their relative contributions to the spin-dependent transport phenomena in the silicon quantum wires has been also established. The fractional form of the $0.7\cdot(2e^2/h)$ feature of the hole quantum conductance staircase has been found by varying the Rashba SOI value.

## 5. Acknowledgement

The work was supported by the programma of SNSF (grant IB7320-110970/1), RAS-QN (grant 4-2. 9A-19), RAS-QM (grant P-03. 4.1).